\documentclass[aps,preprint]{revtex4}%
\usepackage{amsfonts}
\usepackage{amsmath}
\usepackage{amssymb}
\usepackage{graphicx}%
\providecommand{\U}[1]{\protect\rule{.1in}{.1in}}

\begin{document}
\title{A Note Concerning von Neumann Projector Measurement Pointer States, Weak
Values, and Interference }
\author{A. D. Parks and J. E. Gray}
\affiliation{Electromagnetic and Sensor Systems Department Naval Surface Warfare Center
Dahlgren, VA 22448 }
\affiliation{}
\affiliation{}


\begin{abstract}
It is well known that the pointer state $\left\vert \Phi\right\rangle $
resulting from the von Neumann measurement of a projection operator
$\widehat{A}$ performed upon an ensemble of quantum systems in the preselected
state $\left\vert \psi\right\rangle $ depends upon $\widehat{A}\left\vert
\psi\right\rangle $. Here it is shown that the pointer state $\left\vert
\Psi\right\rangle $ obtained from such a measurement performed upon an
ensemble that is also postselected depends upon the weak value of
$\widehat{A}$ - \textit{regardless of the measurement interaction strength}.
It is also found that, while the spatial distribution of $\left\vert
\Psi\right\rangle $ exhibits interference, \textit{the idempotency of
}$\widehat{A}$\textit{ prohibits interference} in that of $\left\vert
\Phi\right\rangle $. This is explained in terms of \textit{welcher Weg} information.

\end{abstract}
\maketitle

\section{Introduction}

The weak value $A_{w}$ of a quantum mechanical observable $A$\ is the
statistical result of a standard measurement procedure performed upon a
preselected and postselected (PPS) ensemble of quantum systems when the
interaction between the measurement apparatus and each system is sufficiently
weak - i.e., when the measurement is a weak measurement \cite{1,2,3}. Unlike a
standard strong measurement of $A$ which significantly disturbs the measured
system and yields the mean value of the associated operator $\hat{A}$ as the
observable's measured value, a weak measurement of $A$ performed upon a PPS
system does not appreciably disturb the quantum system and yields $A_{w}$ as
the observable's measured value. The peculiar nature of the virtually
undisturbed quantum reality that exists between the boundaries defined by the
PPS states is revealed by the eccentric characteristics of $A_{w}$, namely
that $A_{w}$ can be complex valued and that its real part can lie far outside
the eigenvalue spectral limits of $\hat{A}$. While the interpretation of weak
values remains somewhat controversial, experiments have verified several of
the unusual properties predicted by weak value theory \cite{4,5,6,7,8,9,10}.

The impetuses for writing this note are discussions appearing in the recent
literature concerning: (a) the ubiquitous and universal nature of weak values,
\cite{11,11a}; (b) the production of weak values without weak measurements
\cite{12}; and (c) the exact all order theory for weak measurements of
operators $\hat{A}$ which satisfy $\hat{A}^{2}=\hat{1}$ \cite{13}. Here, in
deference to items (a) - (c), the exact pointer theories for arbitrarily
strong von Neumann measurements of both preselected (PS) and PPS systems are
developed for operators satisfying $\hat{A}^{2}=\hat{A}$ (i.e. for
projectors). These theories show that - unlike the pointer states for PS
measurements which depend upon the action $\hat{A}\left\vert \psi\right\rangle
$ of $\hat{A}$ upon the PS state $\left\vert \psi\right\rangle $\ - those for
PPS measurements depend upon $A_{w}$, \textit{regardless of the measurement
interaction strength}; and that interference occurs in the spatial
distribution of a PPS pointer state but is prevented from occurring in the
spatial distribution of a PS pointer state by the idempotency of $\hat{A}$.

\section{Exact Pointer Theories for von Neumann Projector Measurements}

Projection operators are an important part of the general mathematical
formalism of quantum mechanics. There has been a recent increased interest in
these operators because the measurement and interpretation of their weak
values have played a central role in the theoretical and experimental
resolution of foundational issues associated with the quantum box problem and
Hardy's paradox, e.g \cite{6,14}, as well as in the experimental observations
of dynamical non-locality induced effects \cite{15}.

These operators are also interesting because their idempotent property
provides for an exact description of the pointer state resulting from their
measurement. Specifically, when an impulsive von Neumann measurement is
performed upon a quantum system to determine the value of a time independent
projection operator $\widehat{A}$, the associated measurement operator can be
written exactly as%
\begin{equation}
e^{-\frac{i}{\hbar}\gamma\hat{A}\hat{p}}=\hat{1}-\hat{A}+\hat{A}\hat{S},
\label{1}%
\end{equation}
where use has been made of the fact that $\hat{A}^{n}=\hat{A}$, $n\geq1$. Here
$\hat{p}$ is the pointer momentum operator conjugate to the position operator
$\hat{q}$, $\gamma$ is the measurement interaction strength, and $\hat
{S}\equiv e^{-\frac{i}{\hbar}\gamma\hat{p}}$ is the pointer position
translation operator defined by its action $\left\langle q\right\vert \hat
{S}\left\vert \phi\right\rangle \equiv\phi\left(  q-\gamma\right)  $\ upon the
initial pre-measurement pointer state $\left\vert \phi\right\rangle $ (it is
hereafter assumed that $\left[  \hat{A},\hat{S}\right]  =0$).

\subsection{PS Systems}

As a consequence of eq.(\ref{1}), the exact normalized pointer state resulting
from a measurement at time $t$ of a quantum system prepared in the normalized
PS state $\left\vert \psi\right\rangle $ is
\begin{equation}
\left\vert \Phi\right\rangle =e^{-\frac{i}{\hbar}\gamma\hat{A}\hat{p}%
}\left\vert \psi\right\rangle \left\vert \phi\right\rangle =\left(  \hat
{1}-\hat{A}+\hat{A}\hat{S}\right)  \left\vert \psi\right\rangle \left\vert
\phi\right\rangle \label{1c}%
\end{equation}
(the normalization of $\left\vert \Phi\right\rangle $\ follows directly from
the fact that $\left(  \hat{1}-\hat{A}+\hat{A}\hat{S}\right)  ^{-1}=\left(
\hat{1}-\hat{A}+\hat{A}\hat{S}\right)  ^{\dag}=\left(  \hat{1}-\hat{A}+\hat
{A}\hat{S}^{\dag}\right)  $). The associated exact spatial distribution
profile $\left\vert \left\langle q|\Phi\right\rangle \right\vert ^{2}$ of the
pointer is given by%
\begin{equation}
\left\vert \left\langle q|\Phi\right\rangle \right\vert ^{2}=\left(
1-\left\langle \psi\right\vert \hat{A}\left\vert \psi\right\rangle \right)
\left\vert \left\langle q|\Phi\right\rangle \right\vert ^{2}+\left\langle
\psi\right\vert \hat{A}\left\vert \psi\right\rangle \left\vert \left\langle
q\right\vert \hat{S}\left\vert \phi\right\rangle \right\vert ^{2}, \label{1a}%
\end{equation}
and is simply the weighted sum of the distribution profiles for the
pre-measurement state $\left\vert \phi\right\rangle $ and $\hat{S}\left\vert
\phi\right\rangle $ - the pre-measurement state translated by $\gamma$.
Observe that \textit{the idempotency of} $\hat{A}$ \textit{precludes the
existence of an interference cross term} proportional to $\operatorname{Re}%
\left\langle q|\phi\right\rangle ^{\ast}\left\langle q\right\vert \hat
{S}\left\vert \phi\right\rangle $ in eq.(\ref{1a}) because the cross terms
contain $\left\langle \psi\right\vert \hat{A}\left(  \hat{1}-\hat{A}\right)
\left\vert \psi\right\rangle =\left\langle \psi\right\vert \left(  \hat
{A}-\hat{A}^{2}\right)  \left\vert \psi\right\rangle =\left\langle
\psi\right\vert \left(  \hat{A}-\hat{A}\right)  \left\vert \psi\right\rangle
=0$ and $\left\langle \psi\right\vert \left(  \hat{1}-\hat{A}\right)  \hat
{A}\left\vert \psi\right\rangle =0$ as factors.

\subsection{PPS Systems}

Now suppose that a measurement of projector $\hat{A}$\ is performed at time
$t$ upon a PPS system. Then the exact normalized pointer state immediately
following the postselection measurement is given by
\begin{equation}
\left\vert \Psi\right\rangle =\frac{e^{i\chi}}{N}\left(  1-A_{w}+A_{w}\hat
{S}\right)  \left\vert \phi\right\rangle , \label{2}%
\end{equation}
where $\left\vert \psi_{i}\right\rangle $ and $\left\vert \psi_{f}%
\right\rangle $, $\left\langle \psi_{f}|\psi_{i}\right\rangle \neq0$, are the
normalized pre- and postselected states at $t$, respectively; $A_{w}$ is the
weak value of $A$ at $t$ defined by%

\[
A_{w}=\frac{\left\langle \psi_{f}\right\vert \hat{A}\left\vert \psi
_{i}\right\rangle }{\left\langle \psi_{f}|\psi_{i}\right\rangle };
\]
$\chi$ is the Pancharatnam phase defined by \cite{16}%
\[
e^{i\chi}=\frac{\left\langle \psi_{f}|\psi_{i}\right\rangle }{\left\vert
\left\langle \psi_{f}|\psi_{i}\right\rangle \right\vert };
\]
and%
\[
N=\sqrt{\left\vert 1-A_{w}\right\vert ^{2}+\left\vert A_{w}\right\vert
^{2}+2\operatorname{Re}\left[  A_{w}\left(  1-A_{w}^{\ast}\right)
\left\langle \phi\right\vert \hat{S}\left\vert \phi\right\rangle \right]  }.
\]
The exact expression for the pointer's spatial probability distribution
profile is%
\begin{equation}
\left\vert \left\langle q|\Phi\right\rangle \right\vert ^{2}=\left(  \frac
{1}{N^{2}}\right)  \left\{
\begin{array}
[c]{c}%
\left\vert 1-A_{w}\right\vert ^{2}\left\vert \phi\left(  q\right)  \right\vert
^{2}+\left\vert A_{w}\right\vert ^{2}\left\vert \phi\left(  q-\gamma\right)
\right\vert ^{2}+\\
2\operatorname{Re}\left[  A_{w}\left(  1-A_{w}^{\ast}\right)  \phi\left(
q\right)  ^{\ast}\phi\left(  q-\gamma\right)  \right]
\end{array}
\right\}  . \label{2a}%
\end{equation}

The effect of postselection upon pointer states can be seen by comparing
eqs.(\ref{1c}) and (\ref{2}). \textit{Eventhough the measurements are
generally not weak measurements}, it is interesting that - unlike projector
measurement pointer states for PS systems which depend upon $\widehat{A}%
\left\vert \psi\right\rangle $ - \textit{projector measurement pointer states
for PPS systems depend explicitly upon the projector's weak value} $A_{w}$.

Comparison of eqs.(\ref{1a}) and (\ref{2a}) also shows that - in addition to
being a weighted sum of distribution profiles for $\left\vert \phi
\right\rangle $\ and $\hat{S}\left\vert \phi\right\rangle $ - \textit{the
pointer state distribution profile for PPS systems contains interference cross
terms that are induced by state postselection}. Interference occurs here
because postselection nullifies the idempotency of $\hat{A}$ by replacing
$\hat{A}\left\vert \psi\right\rangle $ with $A_{w}$ - thereby allowing the
cross terms to occur. More specifically - unlike a PS measurement where cross
terms contain the vanishing $\left\langle \psi\right\vert \hat{A}\left(
\hat{1}-\hat{A}\right)  \left\vert \psi\right\rangle $ and $\left\langle
\psi\right\vert \left(  \hat{1}-\hat{A}\right)  \hat{A}\left\vert
\psi\right\rangle $ factors - the cross terms for a PPS measurement contain
$A_{w}\left(  1-A_{w}^{\ast}\right)  $ and its complex conjugate as
non-vanishing factors.

\section{Closing Remarks}

The fact that PPS pointer states produced by von Neumann projector
measurements of arbitrary interaction strength depend upon the weak value of
the projector is - perhaps - not surprising in light of the recent discussions
in \cite{11,11a} concerning von Neumann measurements and the associated
ubiquitous and universal nature of weak values. It is also interesting to note
from the comparison of eqs.(\ref{1c}) and (\ref{2}) that PPS pointer states
contain a Pancharatnam phase factor. This is an expected natural consequence
of state postselection \cite{16,17,18}.

Eqs.(\ref{1c}) and (\ref{2}) can also be used to determine additional
differences between the pointers for PS and PPS systems. For example, it is
easy to show that although pointer momentum is not in general a constant of
the motion for von Neumann projector measurements of PPS systems, it is a
constant of the motion for PS systems (in fact this is also true for PS
systems when $\hat{A}$ is not a projector since $\left[  \hat{p},e^{-\frac
{i}{\hbar}\gamma\hat{A}\hat{p}}\right]  =0\Rightarrow$ $\left\langle
\phi\right\vert \left\langle \psi\right\vert e^{\frac{i}{\hbar}\gamma
\widehat{A}\widehat{p}}\hat{p}e^{-\frac{i}{\hbar}\gamma\widehat{A}\widehat{p}%
}\left\vert \psi\right\rangle \left\vert \phi\right\rangle =\left\langle
\phi\right\vert \hat{p}\left\vert \phi\right\rangle $).

Perhaps the most interesting difference revealed by this analysis is related
to interference and can be explained in terms of \textit{welcher Weg}
information. In particular, PS pointer states for projector measurements
contain \textit{welcher Weg} information in the sense that the states
$\left\vert \phi\right\rangle $\ and $\hat{S}\left\vert \phi\right\rangle
$\ that are superposed to form a PS pointer state are "tagged" by the vector
quantities $\left(  \hat{1}-\hat{A}\right)  \left\vert \psi\right\rangle $ and
$\hat{A}\left\vert \psi\right\rangle $, respectively. As shown above, the
idempotency of $\hat{A}$\ naturally precludes the occurrence of PS pointer
state interference. However, postselection effectively "erases" this
\textit{welcher Weg} information by replacing the vector tags with complex
valued weak values of $\hat{A}$\ - thereby enabling PPS pointer states to
exhibit interference.

\section{Acknowledgement}

This research was supported by a grant from the Naval Innovation in Science
and Engineering program sponsored by the Naval Surface Warfare Center Dahlgren Division.

\end{document}